\documentclass[AMA,STIX1COL]{WileyNJD-v2}
\usepackage{moreverb}

\newcommand\BibTeX{{\rmfamily B\kern-.05em \textsc{i\kern-.025em b}\kern-.08em
T\kern-.1667em\lower.7ex\hbox{E}\kern-.125emX}}

\begin{document}
\articletype{Communication}%

%\raggedbottom

\newcommand{\charactercount}[1]{
\immediate\write18{
    expr `texcount -1 -sum -merge #1.tex` + `texcount -1 -sum -merge -char #1.tex` - 1 
    > chars.txt
}\input{chars.txt}}

\title{Direct measurement of current transduction across energy converting cell membranes using a novel membrane-on-a-chip system\protect}

\author[1,3]{Ulrich Ramach}

\author[1]{Rosmarie Sch\"ofbeck}

\author[2]{Jakob Andersson}

\author[1]{Markus Valtiner*}

\authormark{RAMACH \textsc{et al}}

\address[1]{\orgdiv{Institute for Applied Phjysics}, \orgname{Vienna University of Technology}, \orgaddress{\state{Vienna}, \country{Austria}}}

\address[2]{\orgdiv{Biosensor Technologies}, \orgname{Austrian Institute of Technology}, \orgaddress{\state{Tulln}, \country{Austria}}}

\address[3]{\orgdiv{Competence Center for Electrochemical Surface Technologies}, \orgname{CEST}, \orgaddress{\state{Wiener Neustadt}, \country{Austria}}}

\corres{*Markus Valtiner, Vienna University of Technology, Institute for Applied Physics, 1040 Vienna, Austria \email{valtiner@iap.tuwien.ac.at}}

\abstract[Abstract]{The light-driven reactions of photosynthesis, as well as the mitochondrial power supply, are hosted within specialized membranes containing a high fraction of redox-active lipids. Protein mobility and diffusion of redox lipids is believed to be the in plane charge transfer mechanism along such cell membranes. 
Using a \textit{membrane-on-a-chip} setup, we show that redox-active model membranes can conduct and sustain surprisingly high (mA) currents with a specific resistivity typical for semiconductors. 
Our data suggest that charge transfer within cell walls hosting electron-transfer-chains is driven by self-assembling molecular redox-wires that effectively couple redox-proteins by a simultaneous electron and proton \textit{in plane} hopping within a membrane. 
This completely alters our understanding of the role of lipid membranes with wide-range implications suggesting e.g. that conducting membranes may be the precursor for evolving complex redox-machineries of life, and electrochemical membrane deterioration may play an important role in mitochondrial aging. 
Further, these self-assembling organic 2D-conductors offer technologically exploitable features allowing for designing self-assembling and adaptive bio-electronics.}

%\charactercount{main}, 
\maketitle

\title{Self-assembling redox-wires form the 2D power grid of energy converting cell membranes}

\maketitle

\section{Introduction}
\begin{figure}[b]
    \centering
    \includegraphics[scale=0.49]{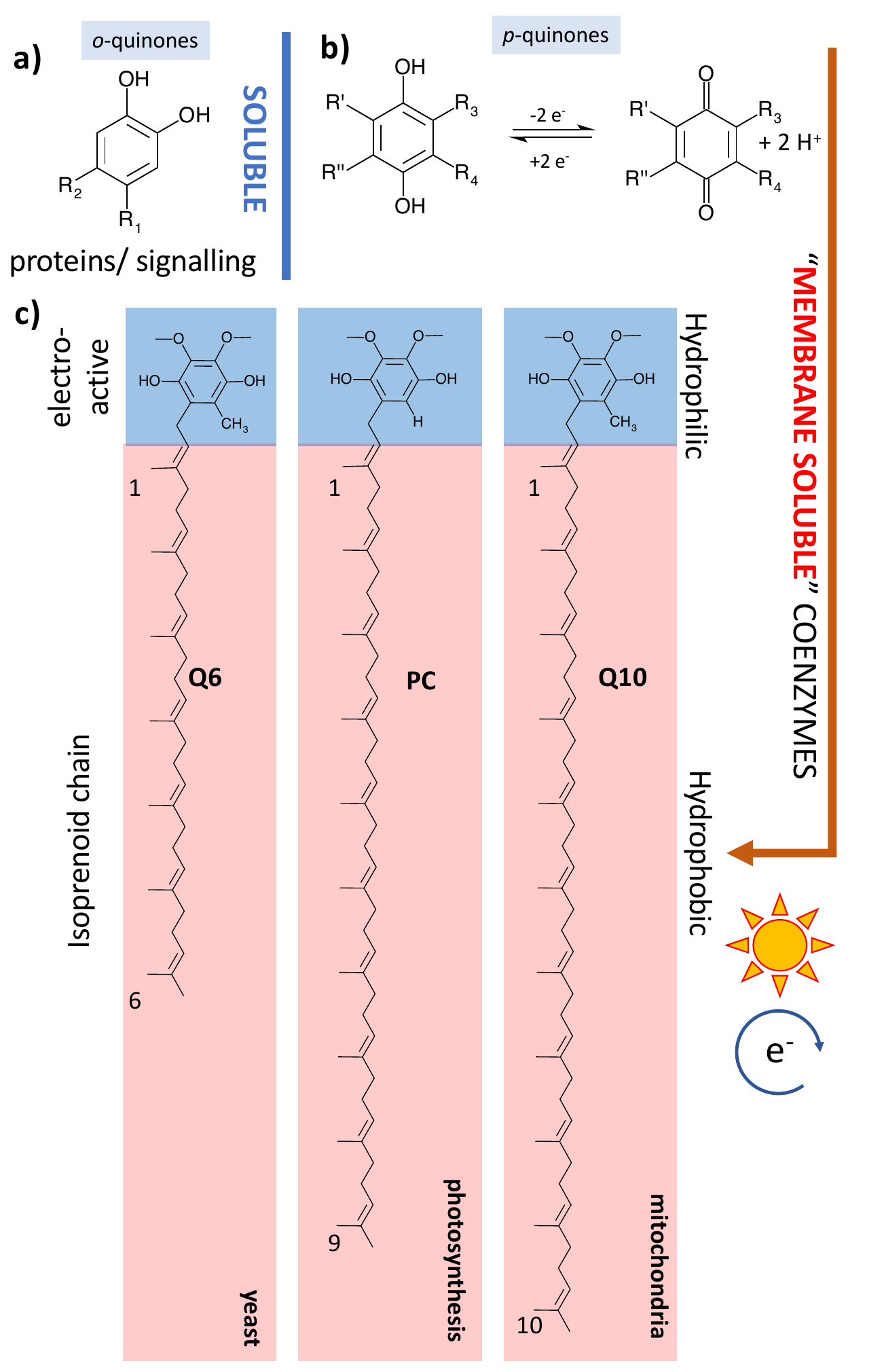}
    \caption
    {
         Overview of natural quinones. 
         a) Water soluble ortho-quinone motives. b) General redox reaction of para-quinones. c) Membrane active para-quinones are linked to hydrophobic isoprenoid chains of varying length (R3). 
    } 
\label{f1}
\end{figure}

Quinone motives in molecules of higher life forms are an abundant molecular electron, redox and proton shuttle, serving very diverse purposes.\cite{Waite2016Sep} For instance, water soluble ortho-quinones (\textbf{Figure \ref{f1}a)} such as adrenaline are important neurotransmitters.
The redox chemistry and pH regulation via the aminoacid L-ortho-Dihydroxyphenylalanin (L-DOPA) plays a central role in wet-glues expressed by mussels. \cite{Guo2020Jan, Waite2002Dec, Lee2011Jul, Waite2016Sep}

Conversely, para-quinones (see \textbf{Figure \ref{f1}b)} act as amphiphilic "membrane-active" redox-shuttles and co-enzymes. For instance, ortho-quinone containing K-vitamins control blood coagulation and support enzymatic activities.\cite{DiNicolantonio2015, Shearer2008Oct} The thylakoid membrane, which hosts the photosynthetic redox-chain, contains plastoquinone (PC) for delivering electrons from photosystem II to plastocyanin, and for building up a proton gradient across a cell membrane.\cite{Junge2019,Zhao2020Jul} 
In mitochondria the oxidative phosphorylation is supported by co-enzyme Q10.\cite{Osellame2012Dec, Friedman2014Jan} 

Species specific Q-enzymes, shown in \textbf{Figure \ref{f1}c}, vary by the number of their isoprenoid side chains. For instance, Q10 is the central electron and proton carrier in human mitochondria, Q6 is active in Saccharomyces cerevisiae.\cite{Parapouli2020} 
\begin{figure}[t]
    \centering
    \includegraphics[scale=0.7]{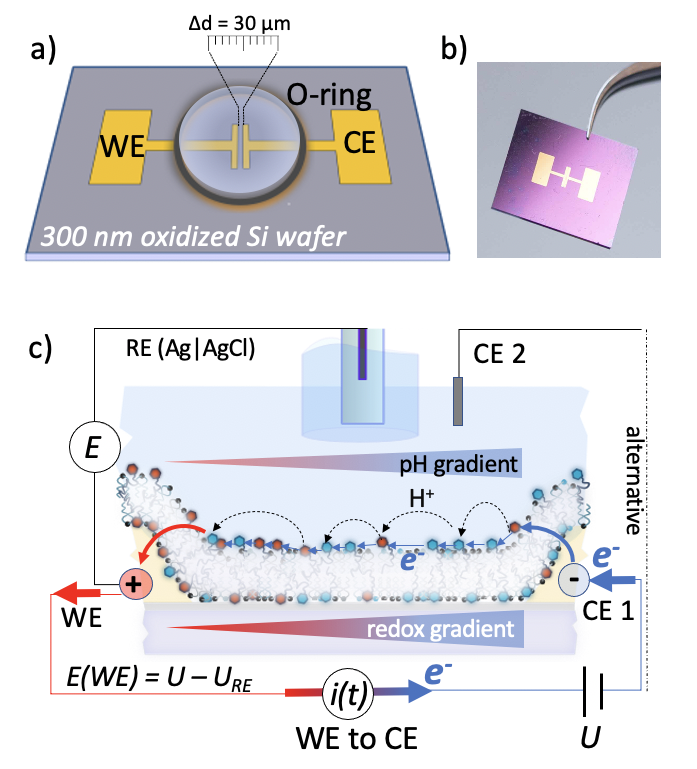}
    \caption
    { 
         a) Schematic of the chip used.  b) Photograph of the chip. c) Schematic of the experimental setup, showing the working (WE), counter (CE1 \textit{or} CE2) and the reference electrode, as well as the depositied lipid bilayer (\textit{c.f.} text for details). 
    } 
\label{f2}
\end{figure}

Hence, in nature membrane soluble quinones appear as para-quinones, with modestly varying chemistry. Redox-membranes are thus electrochemically and conceptually surprisingly similar. 
They feature a high (up to 30\%) content of quinone to provide the reversible redox-equilibrium to sustain electron transport chains (ETC).

It is believed that diffusive electron and proton shuttling via quinones is the major current conduction mechanism along the ETC. 
The observed swelling of the thylakoid membrane under illumination was linked to enhanced mobility and diffusivity of membrane molecules during photosynthesis.\cite{Kirchhoff2011Dec} 
However, diffusion is not rate limiting the ETC from PSII to plastocyanin, but rather the re-oxidation of PC at the plastocyanin redox pocket.\cite{Crofts2013Nov} 
Furthermore, electron bifurcation occurs within Quinone-membranes (Q-membranes) as a central energy minimisation principle for interdependent biologic reaction pathways utilised by nature.\cite{Yuly2020Sep}

Here, we demonstrate electric non-diffusive currents across self-assembled Q-membranes in the mA range.
For this, we designed an electrochemical \textit{membrane-on-a-chip} setup shown in \textbf{Figure \ref{f2} a,b}. 
Two electrodes are placed in close proximity (30 $\mu$m) on the chip. Using a Langmuir Blodgett trough (LBT), we deposited lipid bilayers at controlled surface pressures. As shown in \textbf{Figure \ref{f2}c}, the membrane establishes an \textit{in-plane} connection between the two electrodes, allowing us to measure \textit{in-plane} membrane conductivity and resistance.

\section{Results and Discussions}
For a first set of experiments, we selected a model membrane with 30\%(mol) of Q-enzymes. As matrix, the phospholipid 1,2-Distearoyl-sn-glycero-3-phosphoethanolamine (DSPE) was used which is not redox-active. 
This matrix lipid has an amine headgroup facilitating stable membrane coupling to both materials of the sensor. 
DSPE establishes a lipid bilayer structure with a thickness of about 8 nm,\cite{Israelachvili2011Jun} which can naturally host the hydrophobic tail of Q-enzymes, as shown in the schematic in \textbf{Fig. \ref{f2}c}. 
The formation of stable model membrane architectures with these constituents was tested by LBT and impedance spectroscopy (details see SI, Figures \ref{SI1}-\ref{SI3} and Tables \ref{tab_S1}-\ref{tab_S2}). 

\begin{figure}[t]
    \centering
    \includegraphics[scale=0.6]{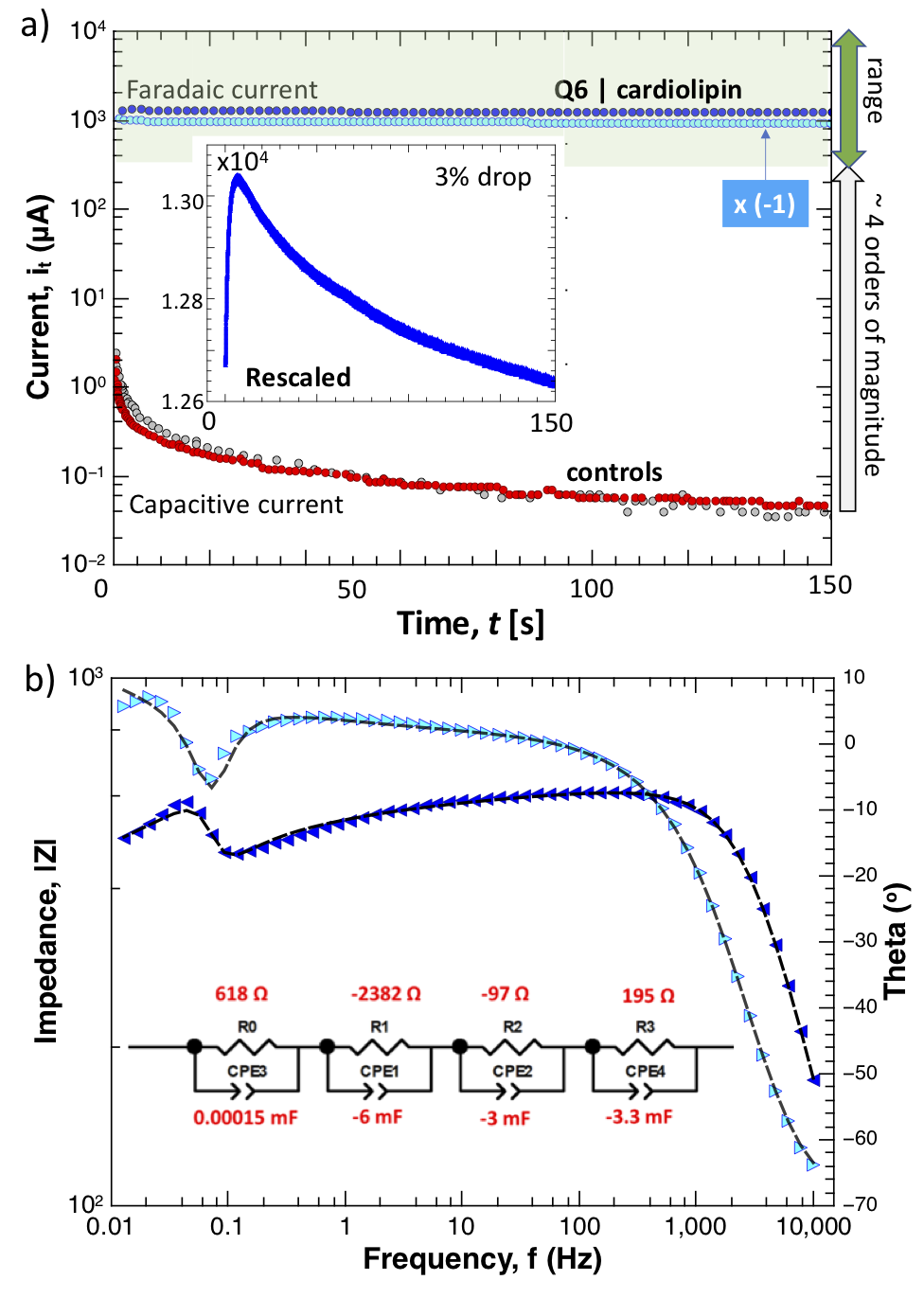}
    \caption
    {
    Electrochemical characteristics of different conductive Q-lipid membranes. a) In plane current between WE and CE1, across 30\% Q6|20\% cardiolipin|DSPE membranes, in comparison to control measurements with DSPE, and blank chips (data points). Exchanging WE and CE1 reverses the current direction, at similar magnitude (see x(-1) data). Data for 30\% Q6|DSPE is shown in the small re-scaled inset, as well indicated by the observed range (arrow on the right). b) Typical impedance spectroscopy of 30\% Q6|20\% cardiolipin|DSPE membranes. Inset: fitting model, with 4 RC-elements in series (\textit{c.f.} text for details). 
    } 
\label{f3}
\end{figure}

As indicated in \textbf{Figure \ref{f2}c}, we adjusted the redox potential of the working electrode (WE) to a potential of +650 mV vs the RE electrode, i.e. above the oxidation potential of the Q-lipid. As indicated in \textbf{Figure \ref{f3}a}, this results in a sustained and high oxidation current ranging from 0.3-14 mA establishes across the membrane. How is it sustained? CE1 simultaneously injects electrons, reduces Q-lipids and the in plane current flow maintains the high level oxidation current at the WE. Hence, an \textit{in plane} electrochemical current flow across the WE and CE is established along the redox gradient.  
The inset in \textbf{Figure \ref{f3}} shows the highest in our experiments measured in-plane current of ~14 mA across 30\% Q6 membranes. Q10 (30\%) was also tested, giving similar current ranges. 

In \textbf{Figure \ref{f3}a}, the measured current is compared to control systems not containing Q-enzymes. Measured controls are the blank chip as well as chips deposited with a 100\% DSPE membrane. 
As expected, the pure DSPE membranes and blank chips exhibit a low and expected capacitive charging of the gold WE in the low $\mu$A range. In contrast, for Q-containing membranes the measured current is $\sim$4 orders of magnitude higher.

In principle, this demonstrates a very effective \textit{in-plane} conduction mechanism across the redox-active membrane. However, with this bilayer chemistry (30\% Q-lipid in a matrix of DSPE), we did not obtain a consistent current with one order of magnitude variation of the current from 0.3-14 mA across different experiments. Also, different patches of Q6/Q10 behaved different. Specifically, we used chemicals as delivered, and in total 4 different patches (>10 sets of experiments each) were tested. From these patches three worked, and one patch did not yield any significant currents in this configuration (30\% Q-enzyme + DSPE). 

We hypothesised that this was due to impurities in the naturally extracted material that can directly effect the mobility, intra-membrane segregation and alignment of Q-lipids. 
Further examination of redox-conducting membranes suggested that intra-membrane segregation of lipid compounds is further moderated by unsaturated lipids. 
For instance, in mitochondria, a special lipid know as cardiolipin\cite{Schlame2000May}, with 4 unsaturated 18 carbon long chains, is present with a concentration of up to 20\%\cite{Schlame2000May}, while thylakoid membranes contain a similar fraction of branched galactolipids (sugar headgroup lipids).\cite{Dormann2002Mar}

We therefore further explored the effect of cardiolipin close to its natural concentration added to our previous membrane design, i.e. 20\% cardiolipin + 30\% Q6 + 50\% DSPE. 
When adding cardiolipin, we find consistent currents of 1-2 mA for all Q6 batches, including the one that did not work well previously, as shown in \textbf{Figure \ref{f3}a}.
In addition, as shown in \textbf{Figure \ref{SI5}}, AFM imaging of the membranes confirms preferential cluster formation and hence pronounced segregation in membranes without cardiolipin. We therefore conclude that the presence of cariolipin prevents clustering. In addition, LBT pressure isotherms indicate a $\sim$35\% increased area per molecule in cardiolipin containing Q-membranes(Figure \ref{SI5}b, compared to membranes without cariolipin (Figure \ref{SI5}a), similarly indicating less compacted \textit{in-plane} membrane structures. 

Hence, membranes where Q-molecules are better dispersed by the unsaturated lipid fraction are essential to ensure the consistent mobility and miscibility to establish stable and reproducible "redox-bridging" across such conducting Q-membranes. 

\textbf{Figure \ref{f3}b} shows a typical in plane potentiostatic impedance spectroscopy of 30\% Q6 | 20\% cardiolipin|DSPE membranes measured at 650 mV applied to the WE, i.e. during current conduction. 

The measured spectral data fit well (red lines) to an equivalent circuit of 4 resistor/capacitor (RC) elements (parameters, see inset) in series \cite{Schonleber2015Sep} with an overall resistance of about 400 Ohm (see \textbf{Figure \ref{f3} b}).
Considering the time domains, the four RC elements approximate 1) the electron transfer resistance at the gold/membrane interfaces (R3/CPE3), while the other 3 elements at lower frequencies may indicate the coupled electron flow (RE1/2 and CPE1/2) and proton translocation (R0/CPE0). 

Interestingly, at lower frequencies we observe impedance behaviour known as an inductive loop\cite{Klotz2019Jan} which is typically found for ion transport in batteries or fuel cells.\cite{Setzler2015Mar, Roy2007Oct} Such features require seemingly nonphysical negative resistances and/or capacitances for equivalent circuit fits. These are often associated with ionic transport in the direction of the potential.
We similarly interpret the low-frequency behaviour as a consequence of in plane transfer of protons, which enhances the electron transfer (redox) reaction. This eliminates the in plane electron transfer resistance, and the overall circuit resistance appears to originate to a large degree from the transfer resistance of the electrode|membrane coupling (R0) and the proton translocation resistance . Diffusion does not play a significant role, as no diffusion element is necessary to fit the data.

This interpretation is in agreement with what we know about biochemical kinetics within ETCs in such membranes. Rate limiting steps are oxidation/reduction of Q-lipids in protein pockets, rather than electron transfer within the Q-lipid layer.\cite{Crofts2013Nov} 

In contrast, the vertical resistances of the membranes measured against CE2 (which cannot short circuit with the WE via the membrane) at 650 mV indicate 200-800 of k$\Omega$ vertical resistances (also \textbf{Figure \ref{SI2}-\ref{SI3}}). 
Further the overall in plane resistance of membranes consisting of DSPE only (control) is in the range of 50 M$\Omega$ ( \textbf{Figure\ref{SI4}}). This resistance can largely be attributed to the conductivity of the hydration layer underneath the membrane.

Using a typical bilayer height of 8 nm\cite{Israelachvili2011Jun} and the electrode width $L = 5 mm$, as well as the electrode distance of 30 $\mu$m, the overall resistance R $\sim$ 400 $\Omega$ translates into a specific resistance of $\rho = R_0\cdot\frac{A}{L}$ = $5\cdot10^{-2}$ [$\Omega$$\cdot$m]. This is lower compared to \textit{e.g.} non-doped silicon and comparable to germanium or GaAs.

Based on the impedance data \textbf{Figure \ref{SI6}} shows a schematic molecular interpretation of the current conduction. Essentially, electrons and protons move in plane along quinone, semi-quinone and hydroxy-quinone in a self-assembling Q-lipid redox chains.  
Conceptually, this is likely similar to bacterial nano-wires\cite{El-Naggar2010Oct, Rowe2018Mar}, where the Fe$^{2+}$|Fe$^{3+}$ redox-couple may transmit current. 

\section{Conclusions}
In summary, fluid redox-active membranes are good and reliable self-assembling conductors based on a proton/electron hoping mechanisms. This suggests that 1) the redox machinery of organic life is "wired up" in 2D. Deterioration of this self-assembling power grid may \textit{e.g.} play an important role in mitochondrial ageing\cite{Sun2016Mar}; 2) that 2D redox-active membranes/liposomes are a potential initial step towards complex ETCs, given that Quinone-membranes are symbiotically acquired machines of higher developed cells,\cite{Lehninger2004Apr} with conserved lipid compositions.\cite{Yoshioka-Nishimura2016Jun} With more complex n-electrodes setups interdependent ETCs can be studied. And 3) that one can exploit such membrane architectures and transient redox-wiring for making self-assembling “living” electronics.

\section*{Acknowledgements}
The authors acknowledge the European Research Council (ERC-StG 663677). 

\section*{Author contributions}
R.S. and U.R. performed experiments. J.A. analyzed impedance measurements, and supported the conceptual development. M.V. conceptually developed this work. 

\section*{Competing Interests}
The Authors declare no competing Interests. 

\section*{Data availability}
All data are available upon reasonable request. 

\bibliography{Library}

\clearpage
 \newpage

\renewcommand{\thefigure}{S\arabic{figure} }
\renewcommand{\thetable}{S\arabic{table} }
\setcounter{page}{1}
\setcounter{figure}{0}
\begin{center}
\begin{huge}
SUPPORTING INFORMATION\\
 \vspace{0.5cm}
Self-assembling redox-wires form the 2D power grid of energyconverting cell membranes\\

\end{huge} 
\vspace{0.5cm}
 {\Large Ulrich Ramach, Rosmarie Sch\"ofbeck, Jakob Andersson and Markus Valtiner}\\
 
\vspace{1cm}
  
\end{center}

\section*{Methods and Materials}
\paragraph{Chemicals and Materials} Phosphate buffered saline (PBS) and other inorganic salts were obtained at p.A. quality (VWR).  
The lipids used were 2,3-dimethoxy-5-methyl-6-(farnesylfarnesyl)-1,4-benzo
quinone (Q6, neat oil, extracted from \textit{saccharomyces cerevisiae}), purchased from Avanti Lipids (via Merck),
Decamethyltetraconta-2,6,10,14,18,22,26,30,34,38-decaenyl]-5,6-dimethoxy-3-methylcyclohexa-2,5-diene-1,4-dione (Q10),purchased from Avanti Lipids (via Merck),
1,2-Distearoyl-sn-glycero-3-phosphoethanolamine (DSPE) as well as 	1',3'-bis[1,2-dioleoyl-sn-glycero-3-phospho]
glycerol (sodium salt) (18:1, cardiolipin) were purchased from Avanti Polar Lipids. The lipids were dissolved in pure chloroform from Carl Roth (assay: $\geq$99.9\%), with lipid mass concentrations of 0.1 mg/ml for Q6 and 0.25 mg/ml for DSPE. In addition Ethanol $\geq$99.9\% pure from VWR and Milli-Q water (Milli-pore, TOC value $\leq$2 ppb, resistivity $\geq$18 M$\Omega$) were used throughout for making soultions and for cleaning equipment. 
The defined volumes of lipid containing solutions of Q6 and DSPE, respectively, were dropped onto the water surface of an LBT (R \& K, Potsdam, Germany) successively.

The lipid containing air/water interface was equilibrated for 10 minutes, and afterwards the lipid mixtures were compressed to 20 mN/m. 
Respective pressure/area characteristics are shown in Fig. S1.
At the set pressure of 20 mN/m the lipid bilayer was deposited by moving a chip out and into the sub-phase of the LBT. After bilayer formation the samples were never allowed to dry out or dewet. Therefore an 8 mm o-ring was glued around the electrodes prior to membrane deposition. 

Milli-Q water (resistivity \textgreater 18 M$\Omega\cdot$ cm, total organic carbon $\leq$4 ppb) was used for making electrolyte solutions. PBS at pH 7 and 10 mM NaCl adjusted to pH 7 were used. 

\paragraph{Chip production}
Chips were prepared by sputtering 5 nm of Ti adhesion layer, and further evaporating 50 nm of gold on an oxidized silicon wafer (100 nm) using a masking technique in a home-build PVD at a base pressure of $2\cdot10^{-6}$. The designed mask generates two gold electrodes that are separated by a 30 µm gap, which was measured by SEM. Using Langmuir-Blodgett trough (LBT) deposition a membrane can be grafted on this chip. An o-ring prevents drying out of the membrane.

\paragraph{Electrochemistry}
All electrochemical experiments were done using either a Pt-wire or Ag|AgCl-electrode as reference electrodes with a PalmSense or a Biologic potentiostatic. Data is referenced to the Ag|AgCl potential. 
A redox potential of +650 mV is established between the working (WE) and reference (RE) electrode, while the counter electrode (CE 1) can supply in plane membrane current. The current is measured laterally across Q-containing membranes as shown in the schematic, which effectively enables an in-plane current between the WE and CE 1. CE 1 supplies the current consumed by the WE in a potentiostatic setup. At the same time a pH gradient is established due to simultaneous proton hoping.  Control experiments include insulating membranes without Q-lipids, blank chips as well as measurements with the redox-active membranes using an alternative counter electrode (CE 2, platinum). CE 2 cannot establish trans membrane current flow as it inserts into the electrolyte without direct membrane contact.

\paragraph{Atomic Force Microscopy}
AFM topography experiments were performed using an environmental Cypher (Asylum Research) and triangular high-frequency cantilevers that are driven by photo thermal excitation. Using a custom-made cell, providing a well for holding liquid, lipid bilayers were deposited and never allowed to dry out or dewet prior to measurements. The well was made by stamping a UV glue well with an o-ring onto muscovite mica sheet where the previously mentioned electrodes were deposited.

\section*{Membrane characterization}
\begin{figure}[b]
\begin{center}
    \includegraphics[scale=0.9]{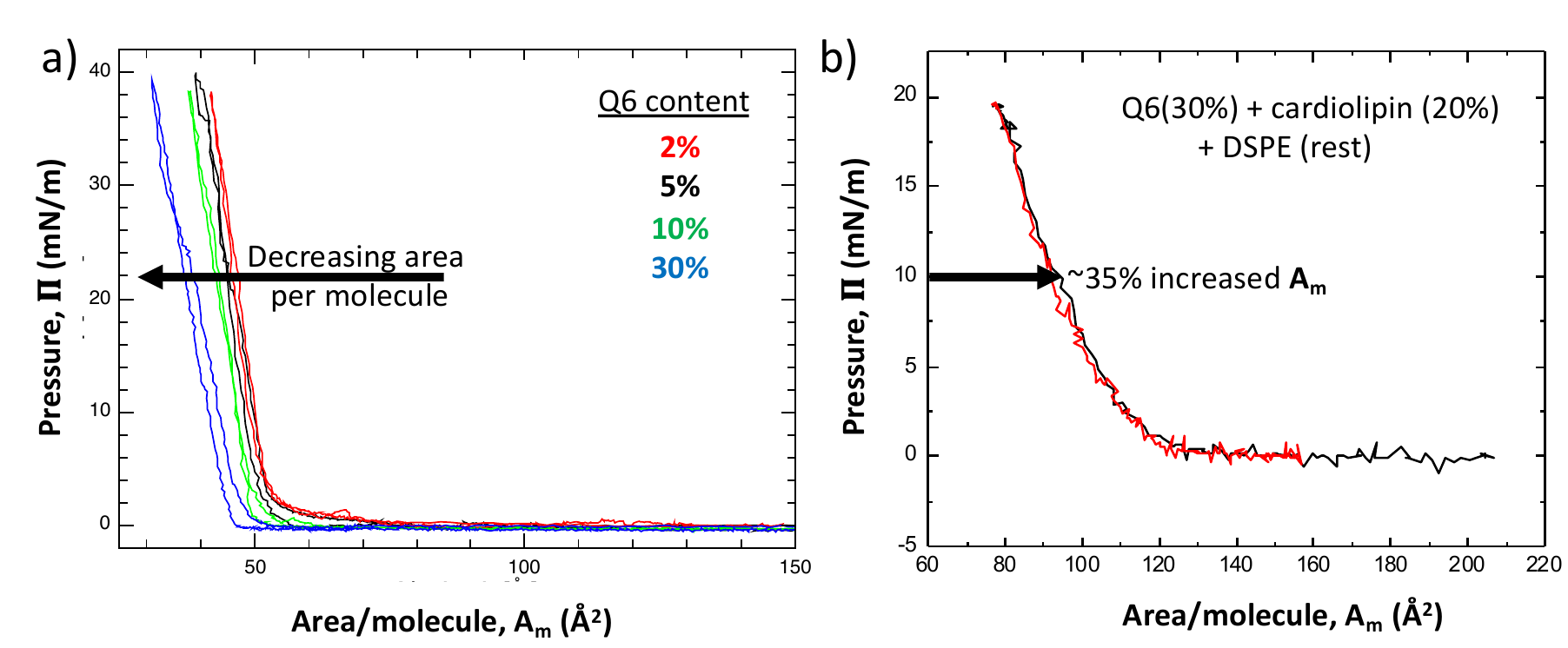}
\end{center}
    \caption
    {
    a) LB pressure area isotherms of DSPE + Q6 at different molar rations. With increasing Q6 content the area per molecule decreases by about 15\% indicating a strong interaction of Q6 and DSPE within the membrane. b) LB pressure area isotherms of DSPE + 30\% Q6 + 20\% cardiolipin. Both systems show a consistent monolayer formation with typical lipid monolayer characteristics. While a) indicates a compression with increasing Q6 content, b) indicates a considerably increased area per molecule with cardiolipin added. As a note: The area per molecule is given as average value over all membrane lipids, and it is not adjusted to linear combinations of the individual constituents. 
    }
    \label{SI1}
\end{figure}

\begin{figure}
\begin{center}
    \includegraphics[scale=0.58]{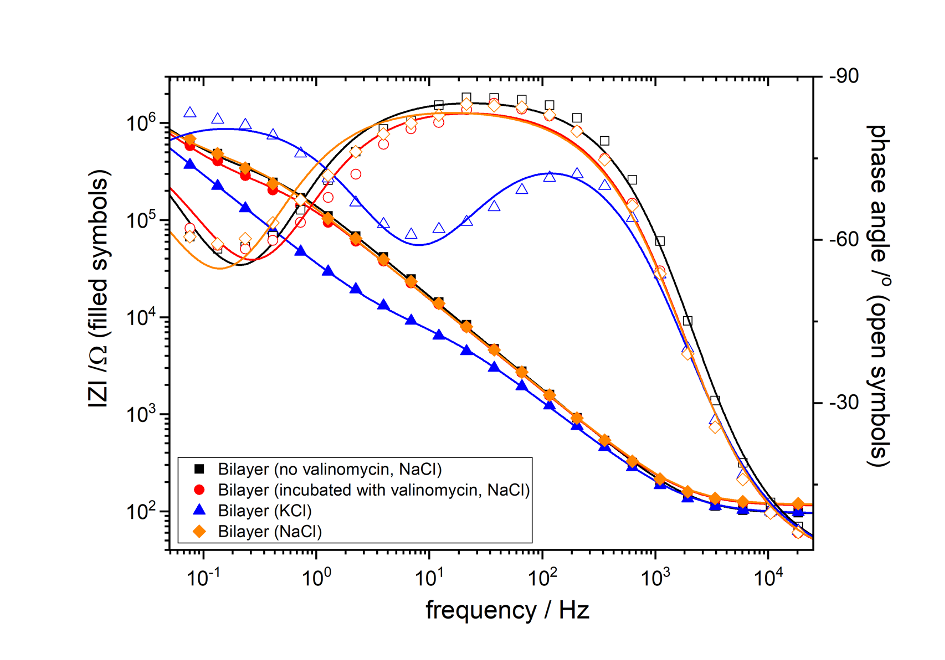}
\end{center}
    \caption
    {
    Bode plots of a coenzyme Q6-modified tethered bilayer lipid membranes functionalised with valinomycin under various electrolytes. See text for details. 
    }
    \label{SI2}
\end{figure}

\begin{figure}
\begin{center}
    \includegraphics[scale=0.58]{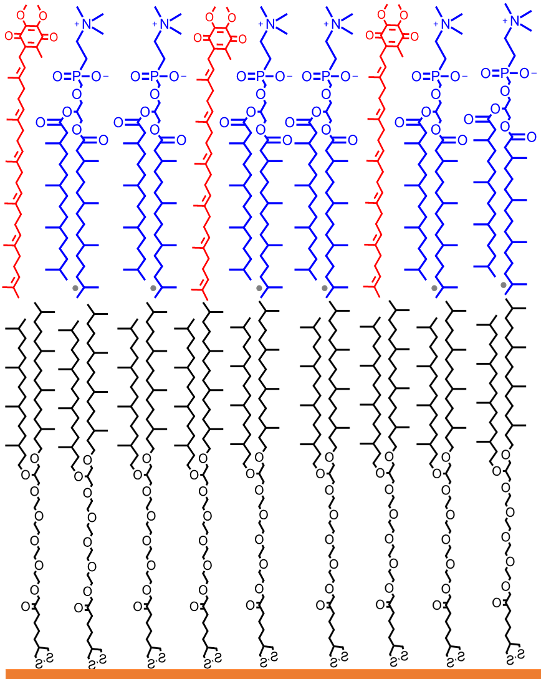}
\end{center}
    \caption
    {
    Schematic of a tBLM functionalised with coenzyme Q6. The inner leaflet is comprised of DPhyTL (black) and the outer leaflet contains a mixture of 1,2-Diphytanoyl-sn-glycero phosphocholine (blue) and coenzyme Q6 (red). 
    }
    \label{SI3}
\end{figure}

Figures \ref{SI2} and \ref{SI3} show impedance of a tethered model membrane with an outer conducting leaflet with and without valinomycin (vmic, an ion conducting channel), proving membrane formation. With valinomycin the resistance can be lowered by adding KCl, which opens the channels, while it increases again with NaCl (see Table \ref{tab_S1}). As shown in Table \ref{tab_S2}, the electrical resistance of  membranes shown in Figures \ref{SI2} and \ref{SI3} is slightly lower than that of one comprised purely of DPhyPC (typically in the range of 10-100 MΩ) and the capacitances are slightly higher than normal. The lower resistances is an indicator that coenzyme Q6 was successfully incorporated. Firstly, because the conductivity of the compound probably amplifies the effect of any defects in the membrane. Secondly, the chain length of coenzyme Q6 does not match that of either DPhyPC or DPhyTL and as such the membrane structure is distorted.

\begin{table}
    \centering
    \caption{Fitted electrical data of a coenzyme Q6-modified tethered bilayer lipid membrane functionalised with valinomycin under various electrolytes.}
    
    \begin{tabular}{|c|c|c|c|c|}
    \hline
      &  Membrane R (k$\Omega$)  &  Error (k$\Omega$) & Membrane capacitance ($\mu$F)  & Error ($\mu$F)   \\
    \hline
       Bilayer  & 183.6  &  19.7 &  1.7 & 0.1  \\
    \hline
       Bilayer after 1h  & 245.8  &  36.3 & 1.6  & 0.1  \\
    \hline
       Bilayer with valinomycin (vmic)  & 161.2  & 22  &  1.9 & 0.1 \\   
    \hline
       Bilayer (vmic) with KCl   &  5.6 &  0.4 &  3.4 &  0.2 \\
    \hline
       Bilayer (vmic) with NaCl   & 294.9  & 50.5  & 1.9  &  0.1  \\
    \hline
    \end{tabular}
    \label{tab_S1}
\end{table}

\begin{table}
    \centering
    \caption{Fitted electrical data of other coenzyme Q6-modified tethered bilayer lipid membrane.}
    \begin{tabular}{|c|c|c|c|c|}
    \hline
         &  Membrane R (k$\Omega$)  &  Error (k$\Omega$) & Membrane capacitance ($\mu$F)  & Error ($\mu$F)  \\
    \hline
       Bilayer 1  & 762.1  &  27.1 &  20.5 & 0.2  \\  
    \hline
       Bilayer 2  & 145.1  &  74.0 & 4.0  & 0.7  \\ 
    \hline
       Bilayer 3  & 185.4  & 0  &  62.6 & 1.1  \\
    \hline
    \end{tabular}
    \label{tab_S2}
\end{table}

\begin{figure}
\begin{center}
    \includegraphics[scale=0.58]{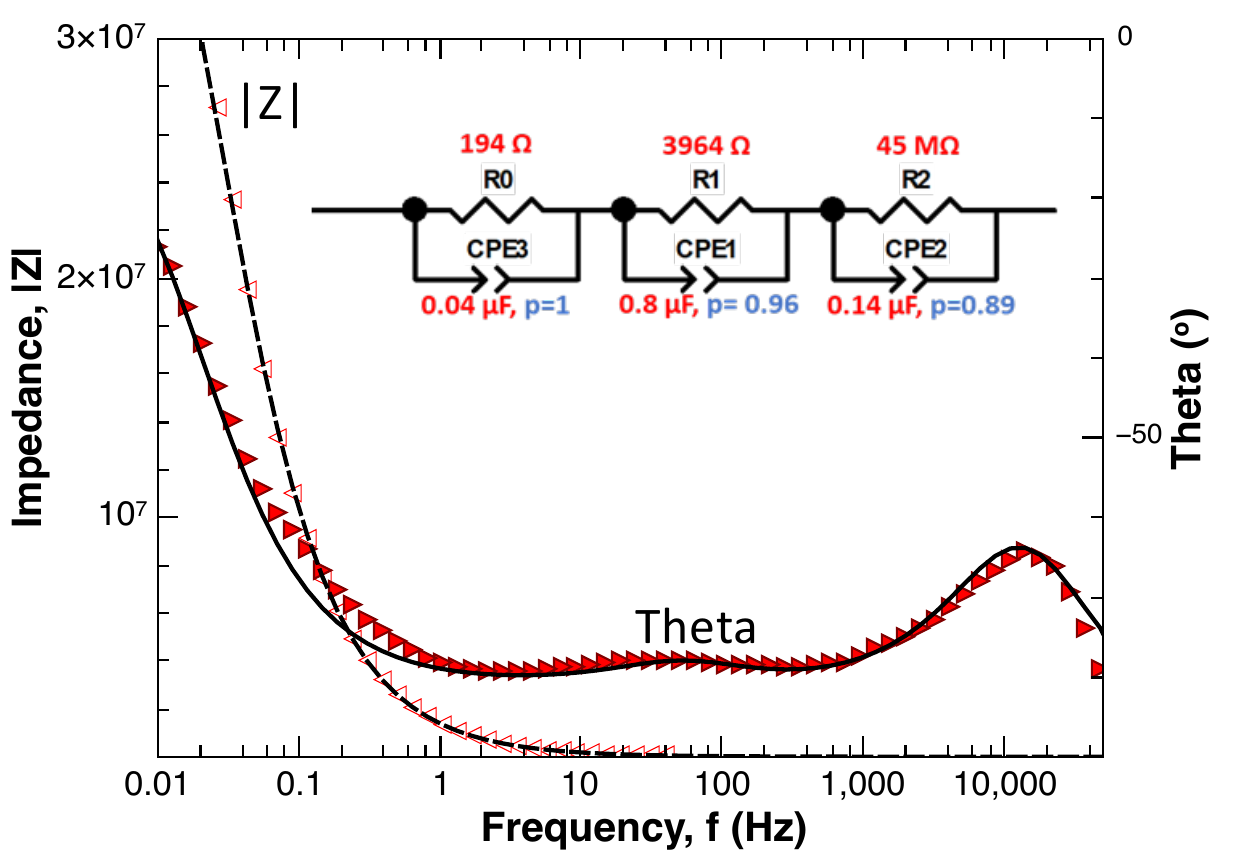}
\end{center}
    \caption
    {
    In plane measured and fitted impedance data of bilayer lipid membranes consisting of DSPE only, measured across WE and CE1. The overall resistance is in the range of 50 M$\Omega$, which is several orders of magnitude larger compared to conducting Q-membranes. 
    }
    \label{SI4}
\end{figure}

\begin{figure}
\begin{center}
    \includegraphics[scale=0.58]{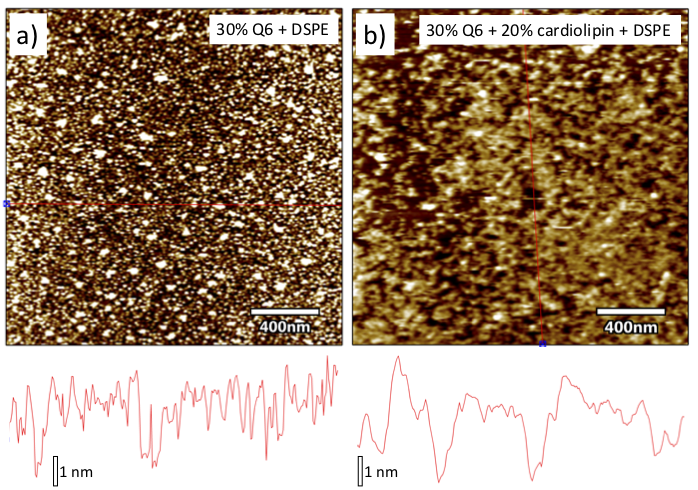}
\end{center}
    \caption
    {
    AFM topography of redox membranes. a) The membrane consisting only of DSPE + 30\% Q6 indicates formation of isolated plaques, indicative of a segregation into the constituents. b) The membrane with an additional 20\% cardiolipin shows no plaque formation, and indicates the formation of a interconnected network of domains.  Also, this layer is very sensitive to pressure during topographic imaging, and at high pressures the outer leaflet can be shaved off leaving a hole that heals with time (data not shown). In contrast, the membrane shown in a) does not heal out after damaging, indicating a high degree of fluidity in the cardiolipin containing membrane.
    }
    \label{SI5}
\end{figure}

\section*{Potential conduction mechanims}
\begin{figure}[h]
    \centering
    \includegraphics[scale=0.75]{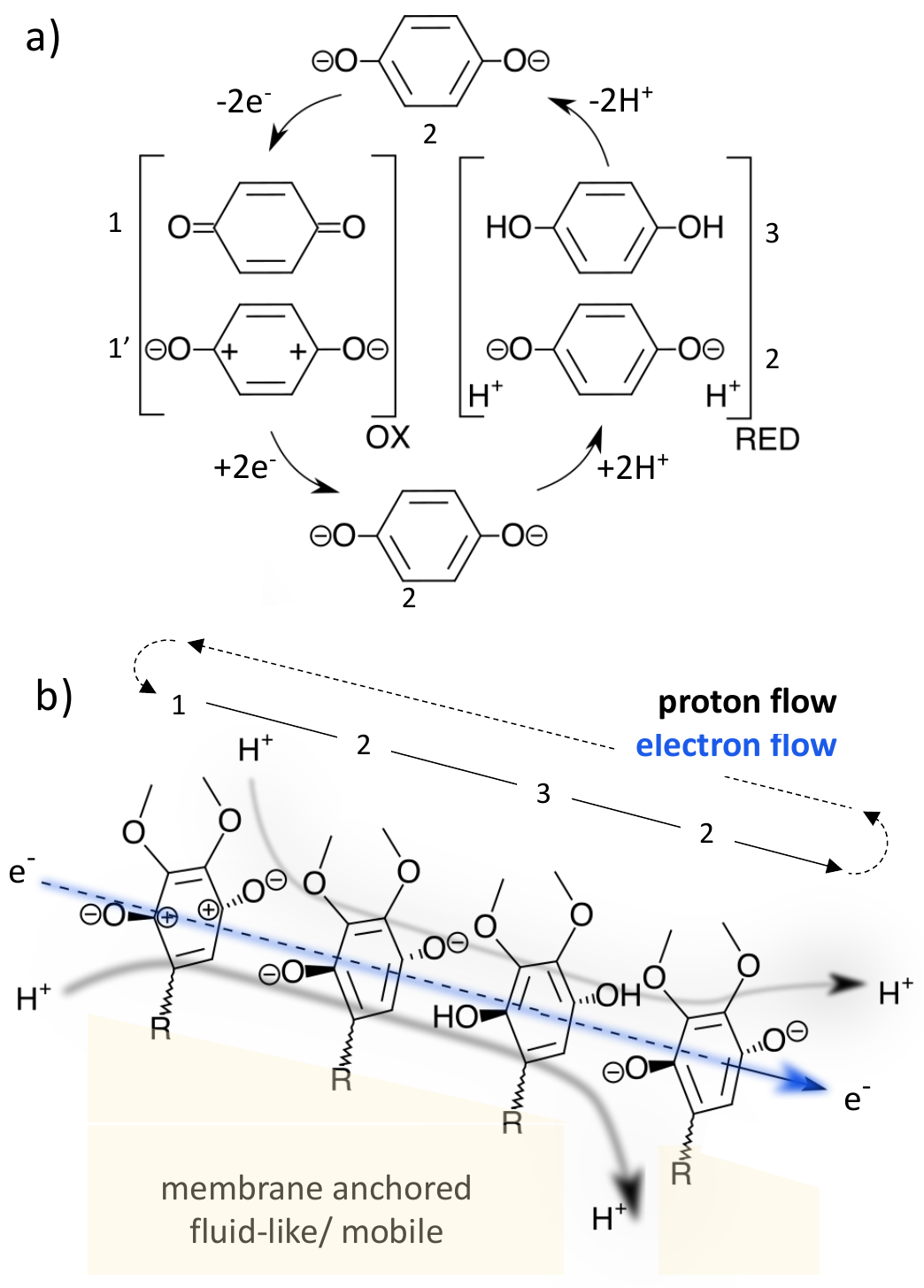}
    \caption
    {a) Redox cycle of a para-quinone. b) Alignment of para-quinones can sustain a continous "rolling" redox cycle enables a directional and simultaneous flow of electrons and protons (\textit{c.f.} text for details).  
    } 
    \label{SI6}
\end{figure}

Based on the impedance data shown in the main text, \textbf{Figure \ref{f3}} shows a schematic molecular interpretation of the suggested current conduction. As indicated in \textbf{Figure \ref{SI6}a}, the oxidized form (1) displayes partial positive charging (1'). The oxidized form can accept electrons when a reduced molecule is in close vicinity, transforming (1') into the reduced but deprotonated form (2). Protons hop with the electrons, yielding the stable reduced form (3). The reduced form again can continue and establish a \textit{rolling cycle} by further pushing electrons to an oxidized form (1) again. Conceptually, this is likely similar to bacterial nano-wires\cite{El-Naggar2010Oct, Rowe2018Mar} where the Fe$^{2+}$|Fe$^{3+}$ couple transmits current. 

\textbf{Figure \ref{SI6} b} shows that such a protonmotive cycle\cite{Mitchell2001Oct} can effectively couple through a chain of para-quinones that are anchored within a bilayer host membrane. The para configuration may allow maximal charge separation of hoping protons, which could be the reason for the general occurrence of p-quinones within respiratory membranes. 
Q-lipids establish a transient wiring, i.e. wiring that can adapt to potential gradients in real time. E.g. switching to negative 650 mV on the WE results in a similar but negative current (see also Figure \ref{f3} a, x(-1)/arrow). 

\end{document}